\documentclass[a4paper, 12pt]{article}

\usepackage[english]{babel}
\usepackage{amsmath}
\usepackage{graphicx}
\usepackage{fullpage}
\usepackage{enumerate}
\usepackage{gensymb}

\title{Security Issues Associated With Error Correction And Privacy Amplification In Quantum Key  Distribution}

\author{ \\Horace P. Yuen\\ \\Department of Electrical Engineering and Computer Science\\Department of Physics and Astronomy\\Northwestern University\\Evanston Il. 60208\\email: yuen@eecs.northwestern.edu}

\date{}

\begin{document}
\linespread{1}
\maketitle
\linespread{2}
\begin{abstract}
          
Privacy amplification is a necessary step in all quantum key distribution protocols, ​and​ error correction is needed in each except when signals of many photons are used in the key communication in quantum noise approach. No security analysis of error correcting code information leak to the attacker has ever been provided, while an ad hoc formula is currently employed to account for such leak in the key generation rate. It is also commonly believed that privacy amplification allows the users to at least establish a short key of arbitrarily close to perfect security. In this paper we show how the lack of rigorous error correction analysis makes the otherwise valid privacy amplification results invalid, and ​that ​there ​exists​ a limit​ ​on how close to perfect a generated key can be​ obtained from privacy amplification. In addition, there is a necessary tradeoff between key rate and security, and the best theoretical values from current theories would not generate enough​ near-uniform​ key bits to cover the message authentication key cost in disturbance-information tradeoff protocols of the BB84 variety​.​

\end{abstract}

\section{Introduction}
Quantum key distribution (QKD) is widely perceived to have provided provably secure key generation with a close to perfect key [1]. The foundational criterion problems that invalidate such conclusion have been described [2,3,4,5]. The security problems associated with error correction and privacy amplification in QKD are briefly discussed in [4,6]. In this paper we focus on these ​latter ​two issues. We will show how intractable the error correction problem is for rigorous analysis, and how that affects the privacy amplification result. We also show the scope and limit of privacy amplification, in particular on​ why​ 
no nearly perfect net key can yet be generated from any known QKD protocol. With the necessary tradeoff between key ra​t​e and security from privacy amplification, this points to the need of developing ​further cryptographic techniques for getting reasonable key rates and security.

\section{Outline of Current QKD Security Analysis}
In this section we summarize the schematic steps in the most complete security analysis
available on concrete QKD protocols of the information-disturbance tradeoff variety, emphasizing the error correction and privacy amplification steps. Such analysis is given for single-photon BB84 including some but not all nonideal effects [7], in particular for no loss and noise in the system. We would describe only the main features of the analysis for the purpose of this paper, and would take as valid the  
steps in [7] not discussed here.

The key $K$ is generated in a QKD round ​through the transmission of random data  $X_j$- carrying optical signals from Alice to Bob, each ​bit $​X_j$ modulating a separate quantum signal. In BB84 type information-disturbance tradeoff protocols, the main focus of this paper, small signals are employed to sense the disturbance from an attacker Eve who taps into the transmission to gain information. The users would use open (public) exchanges to detect Eve's presence. When the quantum bit error rate (QBER) disturbance on a random portion of the signals is above a designed threshold level, the protocol would be aborted. If the QBER is below the threshold, the users through their protocol analysis would estimate that a net key can be generated and proceed as depicted in Fig 1. Before the QBER check, open basis matching ​in each quantum signal ​is carried out between Alice and Bob who measures randomly on one of the bases used in each bit signal. The unchecked portion of $X$ constitutes the sifted key $K''$​, ​on whose carrying ​quantum ​signals Eve has already set her probe attack during transmission. These are all standard BB84 steps.

\begin{figure}[h]
\centering
\includegraphics[width=1\textwidth]{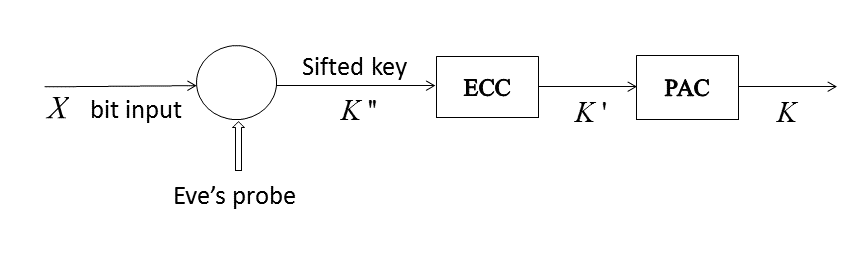}
Fig. 1: Schematic representation of a QKD system incorporating error correction and privacy amplification with generated key $K$.
\end{figure}

The sifted key $K''$ is to be corrected for error, from system imperfection​​ even if Eve is absent. Such system error is sizable for small signals, and very sizable in lossy systems because the received signal rate is greatly reduced in large loss​ as compared to the noise rate​. We ​mainly ​consider the use of error correcting code (ECC)​ ​for such purpose, which in practice is (nearly) universally employed, The ECC output $K'$ is presumably free of error, and is compressed to a shorter key $K$ for increased security by a privacy amplification code (PAC), which in practice is always a matrix compressing the $m$-bit $K'​$​ to an $n$-bit $K$, $m>n$. The final key $K$ is practically now available for use, the key bit cost during protocol execution is subtracted to give the net key bits $|K_g|$ generated. ​T​he open exchanges between the users need to be authenticated with a shared secret key to protect at least against man-in-the-middle attack​, which constitutes the minimum key cost.​ Eve makes measurement on her probe only after she learns about the open information on ECC and PAC.

How does the security analysis go? First, one needs to pick a useful and​ ​workable security criterion. The best criterion to date is the trace distance $d$ [1,8], and we do not need to be concerned with its detailed meaning or adequacy [4,5] in this paper on ECC and PAC​. ​ ​We would just note​ the following.​ 

From her probe measurement result together with whatever side information she has such as the specific ECC or PAC employed, Eve derives a whole probability distribution $P=\{P_i\}$ on the possible values of $K$​ that is objectively correct given the security model is correct​. Let $N=2^n$ and $P_1\geq....\geq P_N$. Thus, $P_1$ is Eve's maximum probability of estimating the whole $K$ correctly during key generation before $K$ is used in an application. It is clear $P_1$ \textit{must be} sufficient small for any security claim. The criterion $d$ is associated with the $K$ generated in one QKD round. It bounds $P_1$ as [4,9,10]
\begin{equation}
P_1\leq d+1/N
\end{equation}
Equality in (1) can be achieved for a given d. ​We c​ompare (1) to $1/N$, the $P_1$ level of a perfect key $U$ (the uniform random variable on $N$ possibilities)​, through the probability exponents​. Let $l$ be the $P_1$ exponent and $n^*$ the $d$ exponent,​ i.e.,​ $P_1=2^{-l}$ and $d=2^{-n^*}$. ​It is clear from (1) that the maximum number of uniform bits that can be extracted from compression on $K$ with a given $d$-level is upper bounded by $n^*$, and with a given $P_1(K)$ ​it ​is upper bounded by $l$.

How is the $d$ level of the PAC output $K$ guaranteed? It comes from the Leftover Hash Lemma, the quantum version of which for $d$ [11] takes the same form as the classical version on  ​
statistical distance [12]. Let $P_1(K')$ be the $P_1$ of the PAC input $K'$ with exponent $l(K')$, $P_1(K')=2^{-l(K')}$. Such $P_1$ exponent is called the minimum entropy $H_{min}$ of the distribution $P$ in the QKD literature. Since $P_1$ has direct operational interpretation, we will ​typically not use ​entropy terminology as the operational meaning of entropic​ ​quantities need to be separately developed​ in general​.   

The Leftover Hash Lemma for ``universal hashing", which covers all PAC in use, gives the tradeoff between the $d$-level $d(K)$ of $K$ and its length $|K|$ by the formula (log to base 2)
\begin{equation}
|K|\geq l(K')-2\log{1/d(K)}\equiv g(K,K')
\end{equation}
​Since it is not known whether $|K|$ bigger than the minimum $g(K,K')$ on the right side of (2) would obtain, the guaranteed key rate is given by the quantity $g(K,K')$. Note that $0\leq d \leq 1$ and thus from (2), $|K|\leq l(K')$. Furthermore, the \textit{minimum} $d(K)$ one can get is, from $0<g(K,K')$,
\begin{equation}
d(K)\geq l(K')^{1/2}
\end{equation}

Actually what are obtained in QKD security proofs are various average results,
say an average $d$-level on $K$ over the universal hashing matrices and average $P_1$ over Eve's measurement results, etc. In this paper we would \textit{not} go into the very significant decrease in quantitative guaranteed level from converting average guarantee to the necessary individual probability guarantee [4,5,13], and
would regard all results as if they are individual.

                                             Thus from (2) the users want as big an exponent $l(K')$, or $H_{min}(K')$ with entropy terminology, as possible. In QKD security proofs, this $l(K')$ is taken to be the \textit{same} as that of the shift key $K'$,
\begin{equation}
l(K') = l(K'')
\end{equation}
To account for the ECC ``information leak" to Eve, the following number of bits $\rm{leak_{EC}}$ is subtracted from $|K|$ to give the net key generated,
\begin{equation}
\rm{leak_{EC}}=f\cdot n\cdot h(\rm{QBER})
\end{equation}
In (5), $h(\cdot)$ is the binary entropy function and $f$ is ad hoc factor usually taken to be $1\leq f\leq 2$. It is not spelled out exactly ​why​ and ​how​ ​(5)​ is the information gain by Eve so that it can be claimed ​that ​the number of net key bits generated is​ (before accounting for message authentication bits),​
\begin{equation}
|K_g|=|K|-\rm{leak_{EC}}
\end{equation}
​The ECC security problems would be discussed in the next section III.​ We would not discuss in this paper the problems associated with bounding $H_{min}(K'')$ under general attacks. Some remarks can be found in [4, 6,14].

                                              So the chain of security analysis for BB84 type information-disturbance tradeoff protocols may be summarized as follows. For a given QBER threshold, a  lower bound on the 
$P_1(K'')$ exponent $l(K'')$, or in entropy notation $H_{min}(K'')$, of the sifted key $K''$ is obtained by some method valid for general attack. The ECC input $K''$ and output $K'$ are taken to have identical $P_1$ exponents, i.e., equ (4) is assumed to hold. Then equ (2) is used to quantify the tradeoff between the PAC output $K$ security level
$d(K)$ and its bit length $|K|$. The PAC side information to Eve is accounted for in the Leftover Hash Lemma (2), and the ECC side information to Eve is accounted for by (5)-(6). There is an $\epsilon$-smooth generalization that improves the tradeoff of (2) which we will discuss in section IV. These steps have been carried out for single-photon BB84 in some detail in [7], but without any specific choice of ECC or PAC. We will go into the ECC and PAC steps separately in the following two sections.​

Before proceeding, it may be mentioned that the information leaks to Eve from error correction and privacy amplification were not accounted for in major previous security analyses such as [15,16]. Despite some recent analyses [17,18] the only available specific evaluation of a concrete protocol is from [7], although it still does not give specific error correction and message authentication procedures. We just see above that error correction and privacy amplification cannot be decoupled in the security analysis in that the output of error correction $K'$ is the input to privacy amplification, and thus
$P_1(K')$ ​and not $P_1(K'')$ ​determines the privacy amplification performance. Similarly, the information leak of $K$ to Eve from message authentication depends on exactly how message authentication is to be carried​ ​out within the QKD protocol with a previously generated imperfect QKD key, which has not been​ described and​ quantified. Thus, we ​would​ ​just ​concentrate in this paper on the specific problems in ​the ​approach of [7], apart from ​these​ incompleteness and other issues.

\section{Security Issues of Error Correction}

 A major problem of QKD protocols to date is that it is very difficult to quantify general security associated with error correction. In the earlier days of QKD, two-way open ``reconciliation" is used to correct errors, specifically through the Cas​c​ade protocol [19]. Cascade is highly nonlinearly random with respect to the various probabilities that occur during its execution. Both the amount of open exchange needed and the leaks to Eve in its various stages are randomly fluctuating. It resists rigorous quantitative analysis similar to
most nonlinear random problems, and various errors of reasoning in its incomplete analysis were ​pointed out in​ [20]. A​ny​ two-way interactive protocol would have this kind of problem. Thus, we would just consider the use of ECC for error correction, which is currently used in concrete protocols.

                                             How is (5) to be justified to be the leak of information to Eve from an ECC?
The case of $f=1$ is supposed to the large $n$ asymptotic limit leak, and $1\leq f\leq 2$ a finite $n$ correction factor. Such choice of $f$ is entirely ad hoc and \textit{out of place} in a rigorous analysis. Leaving $f$ unspecified would make (5) tautologically correct, because any $\rm{leak_{EC}}$ can be given by adjusting $f$. But how does one know a net key can be generated, that (6) is positive for the true $f$? Why is $f$ upper bounded by 2? Why could an imperfect key $K$ at whatever its quantified security level be used for $\rm{leak_{EC}}$ subtraction in (5)-(6)? Such hand-waiving in a claimed rigorous security proof is astounding. In fact, it is shown in [13] that a huge leak from an openly known ECC similar to quantum information locking is possible. In any case, given the size of the ECC required for typical $n$ in the range of $10^4$ and $10^5$​, it appears impossible to quantify any information leak to Eve for an openly known ECC, and no attempt has been reported. Thus, we consider the case of ECC​ that​ is covered up by a shared secret key between the users.

Consider a linear $(m,k)$ ECC with $​$k information digits and $m$ code digits, so that the number of parity check digits is $m-k$ [22]. If the parity check digits of such a linear ECC is covered by shared random bits, the bit cost would be $m-​$k​. If one assumes the QKD system is a binary symmetric channel [10] on the data bits with crossover probability given by QBER, then for $k$ given by the channel bit capacity $n[1- h(\rm{QBER})]$ there exists a (linear) code that would correct the errors from the $n$ received bits by Shannon's channel coding theorem, and hence the number of parity check bits that are to be covered is given by
\begin{equation}
n - n[1-h(\rm{QBER})] = n\cdot h(\rm{QBER})
\end{equation}
This appears to be the argument given in [23] for the $f=1$ case of (5). It seems also to be the argument implicit in [7] ​where​ (5) ​is justified ​by quoting the whole book [10]​, but the book​ does not deal with ``reconciliation" or information leak problems in cryptography.  

There is a minor numerical problem associated with (7), that ​either $|K|=k=n$ or (7) is ​not exactly correct. I​f the users employ ​error free transmission of ​additional​ covered parity check bit​s, $k=n$ and​ $m$ ​becomes​ $n/[1-h(\rm{QBER})$] in the $(m, k)$ code. Thus instead of (5) one ​w​ould have
\begin{equation}
\rm{leak'_{EC}} = n\cdot h(\rm{QBER})/[1 - h(\rm{QBER})]                                   
\end{equation}
for the asymptotic case.​ Generally, there is an additional problem of whether all errors in $K''$ are corrected in $K'$ with a high enough probability. In this paper we assume that is the case​ and also ignore the difference between (5) and (8). ​                                                                           ​
 
In [7] the $f$ of (5) is picked to be 1.1, an ad hoc choice ​that​ greatly affects the number of final net key bits generated given by (6), which is also very sensitive to QBER. The hand-waiving nature of such an estimate is reinforced by several considerations. \textit{First}, there is no reason to expect the QKD system is a memoryless channel under joint attack, especially one with entangled probes by Eve. ​In fact, a general active attack from Eve would invalidate the constant channel assumption
from which the results in [10] are derived​ and [7] refers to​. \textit{Second}, the ECC structural information is not yet accounted for​ ​even if the ECC parity digits are covered by perfect key bits. This is because only a finite amount of such bits can be used as constrained by (6) being positive, and thus the structure of the code has to be openly specified which is also needed for actual implementation. Mathematically, the form of the density operator $\rho(K')$ is changed by such covering from the no-ECC $\rho(K')$ or $\rho(K'')$ to, for $j$th ECC probability $p_j$ that leads to $\rho_j(K'')$ for Eve,
\begin{equation}
\rho '(K')=\sum_j p_j\rho_j(K'')
\end{equation}
​Now the ECC output key $K'$ is in state $\rho'(K')$ to Eve and no longer​ the same as the sifted key state
$\rho(K'')$ before ECC. \textit{Third}, what is the leak when the ECC is covered by a previously generated imperfect key, especially one with ​low ​numerical security level​s​ that can be obtained just theoretically [4,24]?​ ​Note that ``universal composition" does not provide such guarantee [5]. 

Furthermore​,​ from Fig. 1​ one needs a correct bound on $P_1(K')$ of the ECC output $​K'$, which is the PAC input, to guarantee the security level of the final $K$ from (2). If the ECC is open, evidently Eve's $P_1(K')$ in increased from her $P_1(K'')$​ because the ECC information would help Eve to correct her key estimate also​. If ECC is covered, $K''$ is corrected to $K'$ and $P_1(K')$ may ​still ​increase from the known ​ECC ​structural information. Thus, the equality in (4) is generally changed to become an inequality with ``=" replaced by ``$\leq$"​, which is ​less favorable to the users. Certainly the openly known deterministic PAC could only increase $P_1(K​'​​)$​ to $P_1(K)$, and in fact very likely does due to the many-to-one transformation of the PAC​. Thus we have ​the ​chain,
\begin{equation}
P_1(K'') \leq P_1(K') \leq P_1(K)
\end{equation}
Equ (10) immediately shows that the all important $P_1$ security level, and in particular the number of possible (near) uniform bits that can be extracted, gets lower as one performs ECC and PAC. While that is intuitively obvious for ECC, isn't PAC supposed to increase security? Yes that is where the key rate and security tradeoff comes in, as we analyze further in the next section.

\section{Security Issues of Privacy Amplification}
The​ idea of privacy amplification has been well known in random number generation. Its gist is as follows. Consider $m$ independent bits of equal a priori probability 1/2 for the bit values each of which is known to Eve with error probability $p< 1/2$. The mod 2 sum of these $m$ bits is readily shown to be known to Eve with error probability $1/2-\frac{(1-2p)^m}{2}$, which goes to 1/2 for large $m$. ​This $m$ to 1 compression for increasing randomness is inefficient in the sense the random bit generation rate is just $1/m$. Efficiency can be increased if $k$ different linear combinations of the $m$ bits are used to generate $k$ random bits instead. Such matrix compression, or general linear hashing, covers all the practical PAC in current use. 

                                                 There is a security cost to the increased efficiency of $(m, k)$ compression compared to $(m, 1)$ compression. Linear hashing introduces correlation among the $k$ output bits even when the $m$ input bits are independent. In fact, it may produce strong correlations among the output bits for \textit{any}
$m$ input bits, which happens when the compression matrix is degenerate, i.e., having a smaller rank than $k$. The performance of PAC can only be quantified as average over classes of hashing matrices, which always include degenerate matrices [25,26]. If one tests the matrix for degeneracy first before using it as PAC, it is \textit{not} known what the resulting PAC performance would become. So the existence of insecure PAC among a chosen class is a fact, not a mere possibility. It has to be ruled out probabilistically via Markov's inequality [4].

                                                   ​W​hat is the security guarantee for the output $K$ of a PAC with an input $K'$ that has some prescribed characteristics? The \textit{first} main point is that the bits in $K'$ ​are correlated although the data $X_j$ are not. This comes about from at least two sources, the ECC itself and joint attack from Eve. We already just observed above that the bits in $K$ are correlated from the PAC itself. There would be additional correlation that arise from the bit correlation in $K'$. A lot of untutored intuition on assessing the security of $K$ implicitly assumes the bits in $K$ are independent, which is \textit{certainly} not the case. 

The \textit{second} main point is the inequality (10), which says the $P_1$ level can only worsen as the sifted key $K''$ is transformed to the ECC output $K'$ and then the PAC output $K$. Why is then security increased at $K$ compared to $K'$ or $K''$? It is because the security of any given $P_1(X)$ level
of a random variable $X$ is relative to the number of bits in $X$. Let an $n$-bit $K$ has Eve's $P_1(K)$ of the form
\begin{equation}
P_1(K) = 2^{-\lambda n + g(n)},\:\:\:\:\: \rm{lim_{n \to\infty}} \:g(n)/n = 0,  \:\:\:\:0\leq \lambda \leq 1           
\end{equation}
​The minimum $P_1(K)$, the best case from the user's view when $K=U$, occurs when $\lambda=1$ and $g(n)=0$,
i.e., when $P_1(K)=2^{-n}$ which happens if and only if $K$ is perfect. It is clear from (11) that a value of $P_1(K)$, say $2^{-100}$, is perfect for a 100 bits $K$ but not very good for a $10^5$ bits $K$. That is how privacy amplification may give better security, by sacrificing key generation rate.

The impression was widely held from the earlier QKD security proofs that QKD security can be made arbitrarily close to perfect under a fixed key rate below a threshold value
determined from the security analysis, often 10\% or more of $|K''|$. It is supposed to be achievable asymptotically for large $n=|K|$, similar to the original statement of the Shannon Channel Coding Theorem for a memoryless channel [10]. The criterion used is Eve's mutual (accessible) information Iac. It was pointed out since long ago [2,4,27,28] that such conclusion is incorrect. Under an Iac constraint​, 
\begin{equation}
\rm{Iac}(K) ​\leq 2^{-\lambda n+\log{n}} ,\:\:\:\:0 \leq \lambda \leq 1
\end{equation}
it is possible that $P_1(K)$ is as big as [2]
\begin{equation}
P_1(K)​ \sim 2^{-\lambda n}​​                                                             
\end{equation}
While (13) goes to 0 exponentially in $n$, its exponent  $\lambda n$ ​shows it is far from perfect however large $n$ is. It is the \textit{rate} that $P_1(K)$ goes to zero that limits the key rate, not just that it goes to zero. The situation is identical under the $d$ criterion, because under $d(K)=\epsilon$ it is possible that [2,3,4] the bound (1) is achieved with equality. Thus, it is the $d$-exponent $n^*$ which becomes the effective limiting $l(K)$, the exponent of the corresponding $P_1(K)$. As we discuss in connection with (1)-(2), it is the exponent of $P_1(K')$ or of $d(K)$ that determines any tradeoff between security and key rate. The underlying reason for such necessary tradeoff is not that one is dealing with finite and non-asymptotic $n$. It is rather than it is the $P_1$ exponent that determines essential security whatever the number of bits it applies to.

In this connection it is important to recall that the number of bits $n$ in $K$ has to be specified in association with any specific criterion value, be it Iac, $d$, or $P_1$. Failure to do so may engender seriously incorrect conclusion [5].

The \textit{third} main point concerns the limit on the achievable $d$ or $P_1(K)$ that can be obtained for any PAC on given input $K'$ of prescribed security level. It is clear that one cannot always extract all bits given by the exponent $l(K')$ by a fixed known PAC, since the resulting $P_1(K)$ or $d(K)$ level would depend on exactly what Eve's distribution $P(K'$) is. The Leftover Hash Lemma (2) guarantees the output $d(K)$ level only from a single constraint on $P(K')$, in this case $P_1(K')$. It would thus be expected there is a cost for such generality, which is reflected in the tradeoff (2). Similarly, that there is a limit on how small $d(K)$ may be obtained, which we give in (3) above.

                                          It should be mentioned that there is an ``$\epsilon$-smooth" generalization [12] of the tradeoff (2) which may improve the tradeoff quantitatively. Such generalization has not given good numerical values in [7]. In particular, bounds on the numerical values of the $P_1(K')$ exponent can be inferred from the reported values in [7] by assuming the less favorable (2) is used instead. Thus, for one bit generated at $d=10^{-14}$ one can infer the $P_1(K')$ exponent is $l(K')<48$. This corresponds to just 48 uniform key bits, which may not be enough to cover the number of message authentication bits necessary to run the protocol. 
Note that this result is obtained not on individual probabilistic guarantee but with at least two averages   
[4,5] we have mentioned. After the conversion there would just be 16 bits left. Thus, after accounting for message authentication cost there is not net near uniform key bits generated.

                                            Generally, $\epsilon$-smooth generalization is useful for the tradeoff when the $P_1(K)$ level associated with the $d(K)$ level, or equivalently their exponents, is relatively large and one can tradeoff security for key rate increase. The problem of the current situation is, as we see from the above example, that the security level $d(K)$ is \textit{not} adequate even for $\epsilon=0$ and so we would want to decrease the $\epsilon$ in the $\epsilon$-smooth entropy. For small $\epsilon$ there is increasingly less distinction between the original entropy and the $\epsilon$-smooth entropy, which become identical at $\epsilon=0$. Furthermore, since the input $d(K')$ is unknown as it is much harder to bound than $P_1(K'')$, for the relatively large $\epsilon$ used in the $\epsilon$-smooth entropy of [7] it is not clear the PAC has improved any security while it surely decreases the key rate. The $\epsilon$-smooth entropy generalization is interesting theoretically. Its practical importance is yet to be demonstrated.

                                              The PAC security analysis is valid and gives correct conclusion on the output $d(K)$ if the input $P_1(K')$ is correctly bounded. As we observe in section III, there is no such correct bound under the arbitrary procedure (4)-(6), and $P_1(K')$ is bigger than $P_1(K'')$ by an unknown amount. So the final security level $d(K)$ has \textit{not} been validly established although the PAC analysis is unproblematic.

\section{Summary and Conclusion}
The main points ​or problems ​about current ECC and PAC treatments of error correction and privacy amplification are as follows.​

​For ECC:
\begin{enumerate}[(A)]
\item The security treatment of EEC via (4)-(6) ​is unacceptable as rigorous security analysis.
\item Security of ECC appears impossible to quantify. It does not follow from ``universal composition", whether the specific code is covered by a previously generated QKD key or by true uniform bits.
\item The ECC output $P_1(K')$ or minimum entropy $H_{min}(K')$ cannot be rigorously estimated.
\end{enumerate}

Fo​r​ PAC:
\begin{enumerate}[(i)]
\item ​There is an inevitable tradeoff between security level and key rate for asymptotic key length $|K|$ going to infinity also, not just for finite $|K|$.
\item There is a fundamental limit on how small the trace distance criterion may become by privacy amplification.
\item The trace distance security level of the generated key $K$ is not validly derived because the input condition on $P_1(K')$ or $H_{min}(K')$ is not validly derived.​  
\end{enumerate}

 As a consequence, there is no valid security proof in QKD just from point (C) or point (iii) above. The crux of the difficulty lies in a rigorous security treatment of error correction. Note that the approach to QKD security outlined in [29] is fine, except it is mere schematic with no concrete realization. It appears that new techniques would need to be introduced to give correct security proofs for any given mathematical representation. Note that this issue is totally separate from the question of whether the mathematical representation is complete or realistic. The ECC and PAC information leaks are also not taken into account in [30]. Since the security claim of the measurement-device-independent approach [31] to QKD is based on [30] or [15], the security of that approach is not yet established just for these problems alone.

\end{document}